\documentclass[twoside]{dis04}

\def\runauthor{A.~Bacchetta, M.~Radici}
\def\shorttitle{Single-spin asymmetries with 2-hadron fragmentation functions}

\newcommand{\open}{{<\kern -0.3 em{\scriptscriptstyle )}}}
\newcommand{\rr}{|{\bf R}|}

\newcommand{\de}{d}

\begin{document}

\title{SINGLE-SPIN ASYMMETRIES WITH TWO-HADRON FRAGMENTATION FUNCTIONS
}

\author{Alessandro Bacchetta}

\address{Institut f{\"u}r Theoretische Physik, Universit{\"a}t Regensburg,\\
D-93040 Regensburg, Germany}

\author{Marco Radici}

\address{Dipartimento di Fisica Nucleare e Teorica, Universit\`{a} di Pavia, and\\
Istituto Nazionale di Fisica Nucleare, Sezione di Pavia, I-27100 Pavia, Italy}

\maketitle

\abstracts{Using the formalism of two-hadron fragmentation functions, 
  we discuss single-spin asymmetries occurring in the production of two
  hadrons in the current region of deep inelastic scattering,
  with a particular emphasis on transversity
  measurements. 
}

Single-spin asymmetries in single-hadron production have been a subject of
intense activity on the theoretical and experimental sides in the last years,
the reason being that they provide access to the yet unmeasured 
quark transversity distribution, and that they involve interesting effects
related to spin, intrinsic transverse momentum, orbital angular momentum 
and T-odd distribution and fragmentation functions.

Some drawbacks unfortunately affect these observables and hinder the
extraction of clean information on the distribution and fragmentation
functions:
\begin{itemize}
\item{there is no proof of factorization for transverse-momentum dependent
    observables up to subleading twist (only very recently a
    preprint on the leading-twist
    proof appeared~\cite{Ji:2004xq});}
\item{the distribution and fragmentation functions appear in convolutions;}
\item{the evolution equations for transverse momentum dependent functions is
    unknown (only very recently a preprint on this subject appeared~\cite{Idilbi:2004vb});}
\item{the expressions describing the asymmetries have several competing
    contributions.}
\end{itemize}

Two-hadron production asymmetries are free from the first three problems, as
they can be integrated over intrinsic transverse momenta, and 
less affected by the last one. Asymmetries are in this case proportional to
the product of a parton distribution function times a two-hadron fragmentation
function. Single-spin asymmetries contain in particular the so-called
{\em interference fragmentation functions}, which are T-odd, i.e. they are odd
under naive time reversal.

Interference fragmentation functions were studied in Refs.~\cite{Collins:1994kq,Artru:1996zu,Jaffe:1998hf}.
The complete analysis has been carried out up to leading-twist in 
Ref.~\cite{Bianconi:1999cd} and up to subleading-twist in Ref.~\cite{Bacchetta:2003vn}. Positivity bounds and the expansion in the 
partial waves 
of the two hadrons were presented in Ref.~\cite{Bacchetta:2002ux}.
While we refer to this list of references for further details, here we
describe some of the most interesting observables to be measured in
semi-inclusive DIS.

\begin{figure}
\centering
\includegraphics[width=9cm]{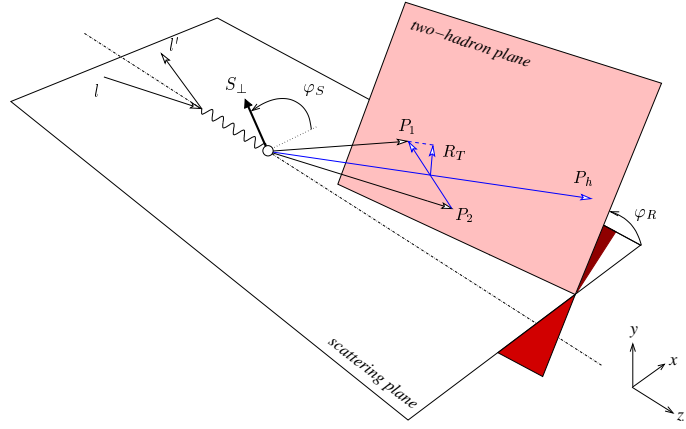}
%\vspace{9cm}
\caption{Description of the angles involved in the measurement of SSA in deep-inelastic
production of two hadrons.}
\label{fig:2hkin}
\end{figure}

The process we are considering is $lp \to l' h_1 h_2 X$, where both hadrons
are produced in the current fragmentation region. The outgoing hadrons have 
momenta $P_{1}$ and $P_{2}$, masses $M_1$ and $M_2$, and invariant mass $M_h$
(which must be much smaller than the virtuality of the photon, $Q$). 
We introduce the vectors $P_h=P_{1}+P_{2}$ and 
$R=(P_{1}-P_{2})/2$, i.e.\ the total and relative momenta of the pair, respectively. The angle 
$\theta$ is the polar angle in the pair's center of mass between the direction of emission 
(which happens to be back-to-back in this frame) and the direction of $P_h$ in any other 
frame~\cite{Bacchetta:2002ux}.
We introduce also the invariant
\begin{equation}
\rr= \frac{1}{2M_h} \, \sqrt{M_h^2 - 2(M_1^2+M_2^2) + (M_1^2-M_2^2)^2}.
\end{equation} 
Cross-sections are assumed to be differential in $\de \cos\theta\,\de
M_h^2\,\de\varphi_R^{}\,\de z\,\de x\,\de y\,\de\varphi_S^{}$, where
$z$, $x$, $y$ are the usual scaling variables employed 
in semi-inclusive DIS and
the azimuthal angles are defined so that (see Fig.~\ref{fig:2hkin})\footnote{Note that there is a
  difference of sign between the angles used here and those used in
  Ref.~\cite{Bacchetta:2003vn}, to conform to the so-called Trento
  conventions.} 
\begin{eqnarray}
  \label{angle-def-1}
\cos \varphi_S &=& 
  \frac{(\hat{\bf q}\times{\bf l})}{|\hat{\bf q}\times{\bf l}|}
  \cdot \frac{(\hat{\bf q}\times{\bf S})}{|\hat{\bf q}
     \times{\bf S}|}, \qquad
\sin \varphi_S \,= \,
  \frac{({\bf l} \times {\bf S}) \cdot \hat{\bf q}}{|\hat{\bf q}
     \times{\bf l}|\,|\hat{\bf q}\times{\bf S}|} , \\
\cos \varphi_R &=&   
%\frac{(\hat{\bf P}_h\times{\bf l})}{|\hat{\bf P}_h\times{\bf l}|}
%  \cdot \frac{(\hat{\bf P}_h\times{\bf R})}{|\hat{\bf P}_h
%     \times{\bf R}|}
%\,\approx\, 
  \frac{(\hat{\bf q}\times{\bf l})}{|\hat{\bf q}\times{\bf l}|}
  \cdot \frac{(\hat{\bf q}\times{\bf R_T})}{|\hat{\bf q}
     \times{\bf R_T}|}, 
%\\
\qquad
\sin \varphi_R 
%&= & 
%  \frac{({\bf l} \times {\bf R}) \cdot \hat{\bf P}_h}{|\hat{\bf P}_h
%     \times{\bf l}|\,|\hat{\bf P}_h\times{\bf R}|}
\,=\, 
  \frac{({\bf l} \times {\bf R_T}) \cdot \hat{\bf q}}{|\hat{\bf q}
     \times{\bf l}|\,|\hat{\bf q}\times{\bf R_T}|} , 
\end{eqnarray}
where $\hat{\bf q} = {\bf q}/|{\bf q}|$ and ${\bf R_T}$ is the component of
$R$ perpendicular to $P_h$.
%and where the approximation sign is
%valid up to subleading twist.

%%%%%%%%%%%%%%%%%%%%%%%%%% Fig. 1 %%%%%%%%%%%%%%%%%%%%%%%%%%%%%
%%%%%%%%%%%%%%%%%%%%%%%%%%%%%%%%%%%%%%%%%%%%%%%%%%%%%%%%%%%%%%%

In writing the following cross sections, it is understood that distribution
functions have a flavor index $a$ and depend on $x$, fragmentation functions
have a flavor index $a$ and depend on $z$, $\cos \theta$ and $M_h^2$. We
introduce the functions
\begin{eqnarray*} 
A(y) &=& 1-y+y^2/2 \; ,
\qquad \qquad
B(y) \,=\, 1-y \; , \\ 
V(y) &=& 2\,(2-y)\, \sqrt{1-y} \; , 
\qquad
W(y) \,=\, 2\, y \,  \sqrt{1-y} \; .
\end{eqnarray*} 
The unpolarized cross section up to subleading twist is 
\begin{eqnarray}
\hspace{-4mm}
\de^7\! \sigma^{}_{UU} &=& \sum_a \frac{\alpha^2 e_a^2}{2\pi Q^2 y}\,\Biggl\{
     A(y) f_1  D_1  
%\\ 
%     & &\qquad \quad 
- V(y) \cos{\varphi_R^{}} \sin\theta \frac{\rr}{Q}  \biggl[ 
     \frac{f_1}{z}\, \widetilde{D}^{\open} + 
     \frac{M x}{M_h}\,  h \, H_1^{\open} \biggr] \Biggr\} .
\label{eq:crossOO}
\end{eqnarray} 
When the target is polarized {\em opposite to the beam direction} the
polarized part of the cross section is
\begin{eqnarray} 
\de^7\! \sigma^{}_{UL'} &=& - \sum_a \frac{\alpha^2 e_a^2}{2\pi Q^2 y}\,\sin{\varphi_{R}}\,\Biggl\{
   |S_L| \, V(y)\, \sin\theta\,\frac{\rr}{Q}\,\biggl[
   \frac{M x}{M_h}\, h_L\, H_1^{\open} +
   \frac{1}{z} \, g_1\,\widetilde{G}^{\open}\biggr]  
\nonumber \\ 
     & & 
   \mbox{}-|{\bf S}_{\perp}^{}|\,B(y)\,\sin\theta\,\frac{\rr}{M_h}\,
   h_1\,H_1^{\open}\; \Biggr\},
\label{eq:crossOL}
\end{eqnarray} 
where $|{\bf S}_{\perp}^{}| = 2\,|S_L|\,M\,x\, \sqrt{1-y}/Q$.
When the target is polarized {\em perpendicular to the beam direction} we have
\begin{eqnarray} 
\hspace{-3mm}
\de^7\! \sigma^{}_{UT} &=& -\sum_a \frac{\alpha^2 e_a^2}{2\pi Q^2 y}\, |{\bf S}_{\perp}^{}|
   \,\Biggl\{ B(y)\, \sin(\varphi_R^{} + \varphi_S^{})\,\sin\theta\,\frac{\rr}{M_h}\,
   h_1\,H_1^{\open} \nonumber \\ 
  & & +V(y)\,\sin{\varphi_S^{}}\,\frac{M_h}{Q}\,\biggl[ h_1\,\biggl( \frac{1}{z}
  \, \widetilde{H} +\sin^2 \theta\,\frac{\rr^2}{M_h^2}\,
   H_1^{\open \, o \,(1)} \biggr) - \frac{M}{M_h} \, x \, f_T^{}
    \, D_1^{} \biggr] \Biggr\} ,
\label{eq:crossOT}
\end{eqnarray} 
When the beam is longitudinally polarized we have
\begin{equation} 
\de^7\! \sigma^{}_{LU} =  -\sum_a  \frac{\alpha^2 e_a^2}{2\pi Q^2 y}\,\lambda_e\,
    W(y)\,\sin{\varphi_{R}}\,\sin\theta\,\frac{\rr}{Q}\,\biggl[
    \frac{M x}{M_h}\, e\, H_1^{\open}
    +\frac{1}{z}\,f_1\,\widetilde{G}^{\open}\biggr] ,
\label{eq:crossLO}
\end{equation}
where $\lambda_e$ denotes the helicity of the lepton.

In Wandzura-Wilzcek approximation, all fragmentation functions with a tilde
vanish.  
Of particular interest is the partial-wave expansion of some of the
fragmentation functions involved, truncated at the $p$-wave level (only $s$
and $p$ waves contribute at low invariant mass):
\begin{eqnarray} 
D_1^{}(z, \cos\theta, M_h^2) &\approx& D_{1,uu}^{}(z, M_h^2) + D_{1,ul}^{}(z, M_h^2)\cos\theta + 
D_{1,ll}^{}(z, M_h^2) {\textstyle \frac{3\cos^2\theta -1}{4}} , \\
\hspace{-3mm}
H_1^{\open}(z, \cos\theta, M_h^2) &\approx&  H_{1,ut}^{\open}(z, M_h^2) + H_{1,lt}^{\open}(z, M_h^2) \cos\theta .
\end{eqnarray} 
The partial-wave expansion shows that, for instance, Eq.~(\ref{eq:crossOT}) 
can be integrated over
$\cos \theta$ without washing out completely 
the term proportional to the transversity
distribution. Unfortunately, 
the dependence on the invariant mass of $H_{1,ut}^{\open}$
is not known, requiring a study in separate invariant-mass bins. Vice-versa,
we expect the term $H_{1,lt}^{\open}$ to show the Breit-Wigner invariant-mass
shape typical of the $\rho$ resonance (for two-pion production): 
in this case   Eq.~(\ref{eq:crossOT}) could be integrated over $M_h^2$
in the neighborhood of the $\rho$ mass, but should be studied in separate
$\cos \theta$ bins in order to disentangle the $H_{1,lt}^{\open}$ contribution.

In conclusion, the measurement of single-spin asymmetries in two-hadron
production in DIS can provide a good way to extract information on the
transversity distribution function $h_1(x)$ and on the distribution function
$e(x)$ in a cleaner way compared to single-hadron production. This kind of
measurements is currently under way at HERMES.

Two-hadron fragmentation functions can be studied also in $e^+e^-$ and $pp$ 
collisions. The first process has been studied in detail in
Ref.~\cite{Boer:2003ya} and it is under experimental study by the BELLE
Collaboration~\cite{Matthias}. The second process can be measured by the
PHENIX and STAR collaborations~\cite{Christine,Renee} and 
allows the measurement of
a convolution of the transversity distribution and the function
$H_1^{\open}$~\cite{Tang:1998wp} 
when employing one polarized proton, and of a convolution of two $H_1^{\open}$
functions when employing two unpolarized protons and detecting two hadron
pairs~\cite{inpreparation}.

\section*{Acknowledgements} 
The work of A.~B. has been supported by the Alexander von Humboldt Foundation.

\end{document}